\documentclass[prx,twocolumn,showpacs,longbibliography,nofootinbib, superscriptaddress]{revtex4-1}

\usepackage{graphics}
\usepackage{graphicx}
\usepackage{amsmath,amssymb}
\usepackage{epstopdf}
\usepackage{color}
\usepackage[normalem]{ulem}

\DeclareGraphicsExtensions{.pdf, .png, .jpg, .eps}
\usepackage{floatrow}

\newcommand{\bfig}{\begin{figure}}
\newcommand{\efig}{\end{figure}}

\newcommand{\kap}{\kappa}

\newcommand{\br}{\boldsymbol{r}}
\newcommand{\bk}{\boldsymbol{k}}

\allowdisplaybreaks

\usepackage{float}

\newcommand{\bea}{\begin{eqnarray}}
\newcommand{\ena}{\end{eqnarray}}
\newcommand{\beq}{\begin{equation}}
\newcommand{\eeq}{\end{equation}}

\newcommand{\hmin}{H_{\rm min}}

\newcommand{\fkap}{\Delta\kappa/\kappa_0}
\newcommand{\fkapH}{\Delta \kappa(H) / \kappa_0}

\newcommand{\black}{\color{black}}

\newcommand{\cybs} {CsYbSe$_{2}$}

\newcommand{\muh} {\mu_0H}

\begin{document}

%\title{Hybridized energy spectrum of the acoustic phonons and spin-flip excitations under magnetic field:  Generic magnetic field dependence of thermal conductivity in the effective spin-1/2 magnetic insulators    }
\title{Heat conduction in magnetic insulators via hybridization of acoustic phonons and spin-flip excitations}

%\title {Generic magnetic field dependence of thermal conductivity in spin-1/2 magnetic insulators enabled by hybrdization of acoustic phonons and spin flip excitations}

%\title { Thermal transport  enabled by hybrdization of acoustic phonons and spin flip excitations in spin-1/2 magnetic insulators enabled by hybrdization of acoustic phonons and spin flip excitations}

%\title{ Universal  magnetic field-dependence of heat conduction in effective spin-1/2 magnetic insulators  by spin-phonon hybridized excitations }

\author{Christopher A. Pocs}
\affiliation{Department of Physics, University of Colorado, Boulder, Colorado 
80309, USA}
%\altaffiliation{Honeywell  Broomfield, CO 80021}

\author{Ian A. Leahy}
\affiliation{Department of Physics, University of Colorado, Boulder, Colorado 
80309, USA}

\author{Jie Xing}
\affiliation{Materials Science and Technology Division, Oak Ridge National 
Laboratory, Oak Ridge, Tennessee 37831, USA}

\author{ Eun Sang Choi}
\affiliation{National High Magnetic Field Laboratory, Tallahassee, Florida, USA}

\author{Athena S. Sefat}
\affiliation{Materials Science and Technology Division, Oak Ridge National 
Laboratory, Oak Ridge, Tennessee 37831, USA}

\author{Michael Hermele}
\affiliation{Department of Physics, University of Colorado, Boulder, Colorado 
80309, USA}
\affiliation{Center for Theory of Quantum Matter, University of Colorado, 
Boulder, Colorado 80309, USA}

\author{Minhyea Lee}
%\email{minhyea.lee@colorado.edu}
\affiliation{Department of Physics, University of Colorado, Boulder, Colorado 
80309, USA}

\date{\today}

\begin{abstract}

%Magnetic insulators provide excellent playgrounds to realize a range of exciting spin models, some of which predict exotic spin ground states, and thermal transport properties have been taking center stage in probing the spin excitations. 
%Despite the fact that acoustic phonons make the major contribution to heat conduction in a crystalline system,  their interplay with magnetic excitations is often viewed as peripheral to the physics of interest, for instance as an inconvenient source of scattering or decoherence.  
%
%Here,
We present a comprehensive study on the longitudinal magneto-thermal transport in a paramagnetic effective spin-1/2 magnetic insulator \cybs, by 
introducing  a minimal model requiring only Zeeman splitting and magnetoelastic coupling. 
We use it to argue that hybridized excitations -- formed from acoustic phonons and 
localized spin-flip-excitations across the Zeeman gap of the crystal electric field ground doublet -- are responsible for a non-monotonic field dependence of longitudinal thermal conductivity.  
Beyond highlighting a starring role for phonons, our results raise the prospect of universal magneto-thermal transport phenomena in %\red 
paramagnetic  insulators that originate from simple features shared across many systems.   \black

\end{abstract}

\maketitle

%and thermal transport properties have been taking center stage in probing the spin excitations. 
%Despite the fact that acoustic phonons make the major contribution to heat conduction in a crystalline system,  their interplay with magnetic excitations is often viewed as peripheral to the physics of interest, for instance as an inconvenient source of scattering or decoherence.  

Magnetic insulators provide excellent playgrounds to realize a range of exciting spin models, some of which predict exotic spin ground states. 
Thanks to its exclusive sensitivity to itinerant excitations, thermal conductivity is one of the most valuable probes for examining magnetic insulators and for  characterizing the magnetic ground state \cite{npjQSLreview2019,  Sologubenko2000, Hess2001, SYLi2005, SavaryReview2016}.  
The magnetic field ($H$) dependence of thermal transport has been argued to be a signature of unconventional spin excitations   \cite{Phuan_TbTiO227_2015,Watanabe2016,Tokiwa2017, Kasahara2018, Akazawa2020, XHong2021, Barthelemy2023}.  However, even though phonon excitations of the crystalline lattice are the dominant heat carriers, much of our understanding relies on viewing them in a subsidiary role. When phonons are not entirely neglected in studying the field-dependence of thermal conductivity, one typically considers a thermal current of phonons scattering off spin excitations. While such perspectives can be useful, they can also be oversimplifications that lead us to miss essential physics.

A case in point is the non-monotonic field dependence of thermal conductivity ($\kap$) that has been observed in the paramagnetic states of several effective spin-1/2 magnetic insulators. Namely,  (1) $\kap(H)$ first decreases to a minimum at $H=\hmin$ followed by an increase and (2) $\hmin$ moves toward large values with increasing temperature.   A handful of systems  have shown  this behavior including Cu$_3$VO$_7$(OH)$_2\cdot$ H$_2$O \cite{Watanabe2016},  YbTiO$_7$ \cite{Tokiwa2017},  Cd-kapellasite  \cite{Akazawa2020}, gadolinium gallium garnet \cite{Tsui1999},  and   $\alpha$-RuCl$_3$ above its ordering temperature \cite{Leahy2017}. %  \textcolor{red}{Universal behavior can reasonably be expected across a range of systems within the paramagnetic state, above temperature scales associated with magnetic ordering and spin-spin exchange interactions.}  
Explaining this phenomenology in terms of spin-phonon scattering alone is not plausible, without invoking unusual responses of the spin sector to applied field that seem unlikely to be present across such a wide range of systems.  Instead, a generic explanation is called for that involves only common ingredients,  that treats phonon and spin excitations on equal footing, %\textcolor{red}{
and that focuses on the paramagnetic state above magnetic ordering  and spin exchange temperature scales, where universal behavior can reasonably be expected.
%}

Here, we propose such an explanation via our study of the well-characterized Kramers pseudospin-1/2 Yb- based triangular lattice \cybs, where we observed the non-monotonic field dependence of $\kap$ described above. We propose that the longitudinal heat conduction under field is enabled by the hybridized quasiparticles formed from acoustic phonons and spin-flip excitations (SFEs)  across the Zeeman gap, and hypothesize that the resulting hybridized excitations are responsible for the non-monotonic $\kap(H)$.  This aligns with a highly simplified theoretical model that qualitatively reproduces key features of the experimental data. 
The two main ingredients of our model, namely (1) single-ion Zeeman splitting and associated SFEs and (2)  magnetoelastic (ME) coupling that mediates phonon-SFE hybridization, are common in magnetic insulators.  Therefore,  our results offer a starting point to understand the non-monotonic field dependence of $\kappa$ observed in  a range of other systems.
%\red
A  quantitative theory of this field dependence will require microscopic calculations able to incorporate detailed characteristics of specific materials, and is a topic for future work.
\black

Hybridization is an archetypical manifestation of quantum mechanics, and there is a long history of studying hybridization between phonons and different types of magnetic excitations \cite{MagnetoelasticBookchapter}, largely with a focus on spectroscopic properties.  
More recently, theoretical works \cite{XZhang2019,Go2019,SPark2019,BMa2022,BMa2023} and one experimental study \cite{NLi2023} have considered magnon-phonon hybridization in magnetically ordered systems, and the resulting Berry phase, as a mechanism to generate a (transverse) thermal Hall signal in insulators.  
In contrast with these prior works, we find dramatic effects of hybridization on the easily measurable longitudinal thermal conductivity.  Moreover, while the previous works on thermal Hall effect require particular magnetic orderings or specific forms of spin-spin interaction, our model relies only on ingredients present in \emph{any} effective spin-1/2 system, namely Zeeman splitting and magnetoelastic coupling.

%  Previous works have studied hybridization. theoretically \cite{Kjems1975, BQLiu2018,XZhang2019,SPark2019,BMa2023} and experimentally \cite{Cermak2019, Ozerov2022, ,NLi2023}  between different types of magnetic excitations and phonons.  While these works  require particular magnetic orderings or specific energy hierachy, our model  includes only   generic standard ingredients for effective spin-1/2 systems regardless of ordering state -- Zeeman splitting and small to moderate magnetoelastic coupling that are applied for  readily accessible longitudinal thermal transport quantities
%%We remark that previous works have studied hybridization between different types of magnetic excitations and phonons \cite{Kjems1975,BQLiu2018,Cermak2019,TKim2019,Ozerov2022}, although their focus has been on spectroscopic probes, and we are not aware of prior studies considering the effects on transport \cite{XZhang2019,SPark2019,BMa2022, NLi2023}. 
%**Ref.\cite{NLi2023} is  experimental paper  arguing the berry phase from maganon-polaron hybridization responsible for the $\kappa_{xy}$ 
% 

 %%%%%%%%%%%%%%%%%%%%%%%%%%%%
%%Figure 1 %%%%%%%Figure 1%%%%Figure 1%%
%%%%%%%%%%%%%%%%%%%%%%%%%%%%

\begin{figure}%[!b]%[ht]
\begin{center}

\includegraphics[width= 1.0\linewidth]{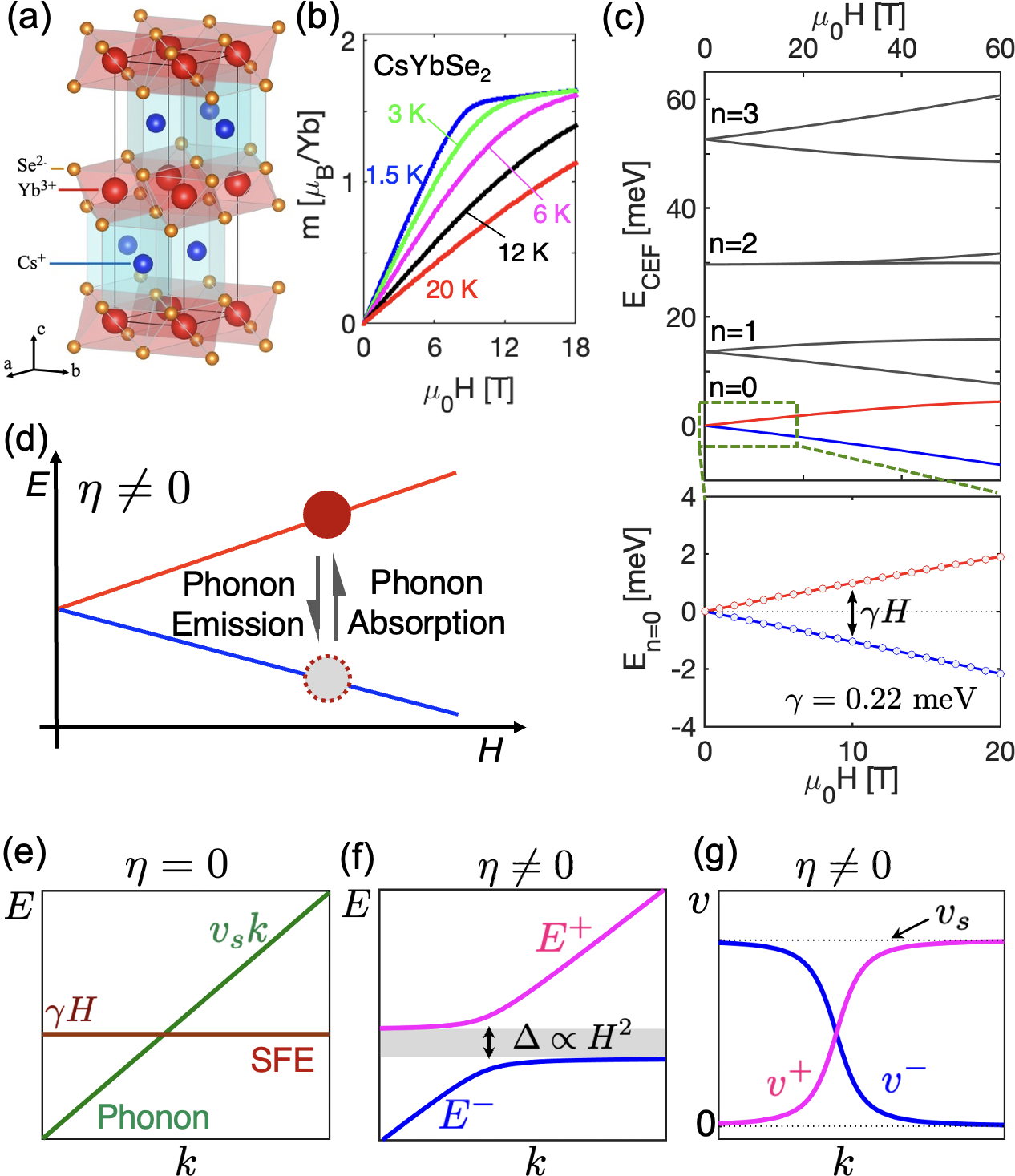}
\caption{ (a) Crystal structure of \cybs: Yb$^{3+}$ (red)  forms triangular nets within the $ab$ plane,  formed by edge-sharing YbSe$_6$ octahedra.  (b) Calculated %\red 
paramagnetic \black single-ion magnetization within Weiss mean-field approximation \cite{Pocs2021}.
(c) Field dependent crystal electric field energy spectrum calculated with the parameters  in ~\cite{Pocs2021}, where  linear splitting of the ground state doublets is shown. 
(d) Schematic illustration of the processes leading to hybridization of phonons and SFEs, where a SFE decays (is created) by emitting (absorbing) a phonon.
%Schematic illustration of  YbSe$_6$  distortion  induced by phonon, which results in the $g$-factor variation   leading to the magnetoelastic coupling strength $\eta$ via the cooperative Jahn-Teller effect \cite{Gehring1975}. 
% {\bf d} Both Zeeman-split ground state branches   (solid lines) and resulting Zeeman gap ($\Delta_Z(H)$, dotted line) as a function of field.
(e) Schematic sketch of the dispersion relation with no ME coupling ($\eta = 0$).  A flat band of SFEs has energy $\gamma H$, independent of wave vector $k$, while an acoustic phonon has a linear dispersion with slope $v_s$. 
(f) Schematic dispersion relation in the presence of non-zero ME coupling ($\eta \neq 0$), where hybridization leads to an avoided crossing and the opening of a gap between upper and lower branches of hybrid SFE-phonon excitations.
(g) Group velocities  $v_\pm=\frac{dE_{\pm}}{dk}$, derived from the dispersion of the hybridized excitations.}
\label{schem}
\end{center}
\end{figure}

 The Yb-based triangular-lattice compound CsYbSe$_2$ has space group $P6_3/mmc$  and consists of layers of edge-sharing YbSe$_6$ octahedra separated by Cs$^{3+}$ ions as shown in Fig.~\ref{schem}(a).  
 Millimeter-sized hexagonal shape CsYbSe$_2$ single crystals were grown by the salt flux method
following the procedure described in Ref. \cite{Xing2020acs}.
The in-plane longitudinal thermal conductivity  ($\nabla T \parallel ab$) was measured on the samples of typical dimensions $1.5 \times 3\times 0.02$ mm$^3$ using a single-heater, two-thermometer configuration in steady-state operation with the field applied in the $ab$ plane parallel to the thermal gradient. 
All thermometry was performed using Cernox resistors, which were precalibrated individually and \emph{in situ} under the maximum applied fields of two different cryostats with superconducting magnets.

 Yb$^{3+}$ carries a $J = 7/2$ magnetic moment, which is split into four doubly degenerate crystal electric field (CEF) levels at zero field.  Single-ion CEF parameters have been determined in \cite{Pocs2021}, 
 allowing for calculations of the %\red 
 paramagnetic \black magnetization 
 within the Weiss-mean field approximation [Fig.~\ref{schem}(b)], which agree well with the magnetization data up to $\muh = 7$ T. The CEF energy spectrum under field was also obtained and shown in Fig.~\ref{schem}(c), where the Kramers doublets are split via the Zeeman effect under applied field. 
%\blue
When temperature ($T$) is lowered below the first excitation energy gap ($\Delta_{10} \simeq 13$ meV), restriction of the dynamics to the ground doublet justifies use of a pseudospin-1/2 model ${\mathcal H}_Z = \gamma H \hat{S}^x$,  where $\hat{S}^i$ are the pseudospin-1/2 operators, we always apply the field along the $x$-axis (within the crystalline $ab$-plane), and $\gamma = 0.22 \, \mathrm{meV}/\mathrm{T}$ \cite{Pocs2021}.  
\black

 %\green[This paragraph can be shortened only consider $g_\perp$ term]  The first order Zeeman splitting of the ground doublet  in \green  the in-plane configruation
 %is encoded in the Hamiltonian 
 %${\mathcal H}_Z = \mu_B \mu_0 ( g_z H_z \hat{S}^z +  g_\perp H_x \hat{S}^x + g_\perp H_y \hat{S}^y)$, 
% where $\hat{S}^i$ are the pseudospin-1/2 operators and $g_z$ and $g_\perp$ the components of 
 %the anisotropic $g$-factor \cite{Pocs2021}. 
%We always consider field applied along the $x$-axis (within the crystalline $ab$-plane), so the Zeeman gap of the ground doublet is $\gamma H= g \mu_B \mu_0 H$, where $g \equiv g_\perp = 2g_J | \langle 0_{\pm}|\hat {J_x}|0_{\mp}\rangle |$.  Here $\hat {J}_x$ is the angular momentum operator in the direction of the applied field,  $| 0_{\pm} \rangle$ are the energy eigenstates of the doublet under applied $z$-axis field, and $g_J$ is the Land\'{e} $g$-factor for the Yb$^{3+}$ ion.
%\black

%%%%%%%%%%%%%%%%%%%%%%%%%%%%
%%Figure  2 %%%%%%%Figure 2  %%%%Figure 2%%
%%%%%%%%%%%%%%%%%%%%%%%%%%%%
\begin{figure}[ht]
\begin{center}
\includegraphics[width= 1.0\linewidth]{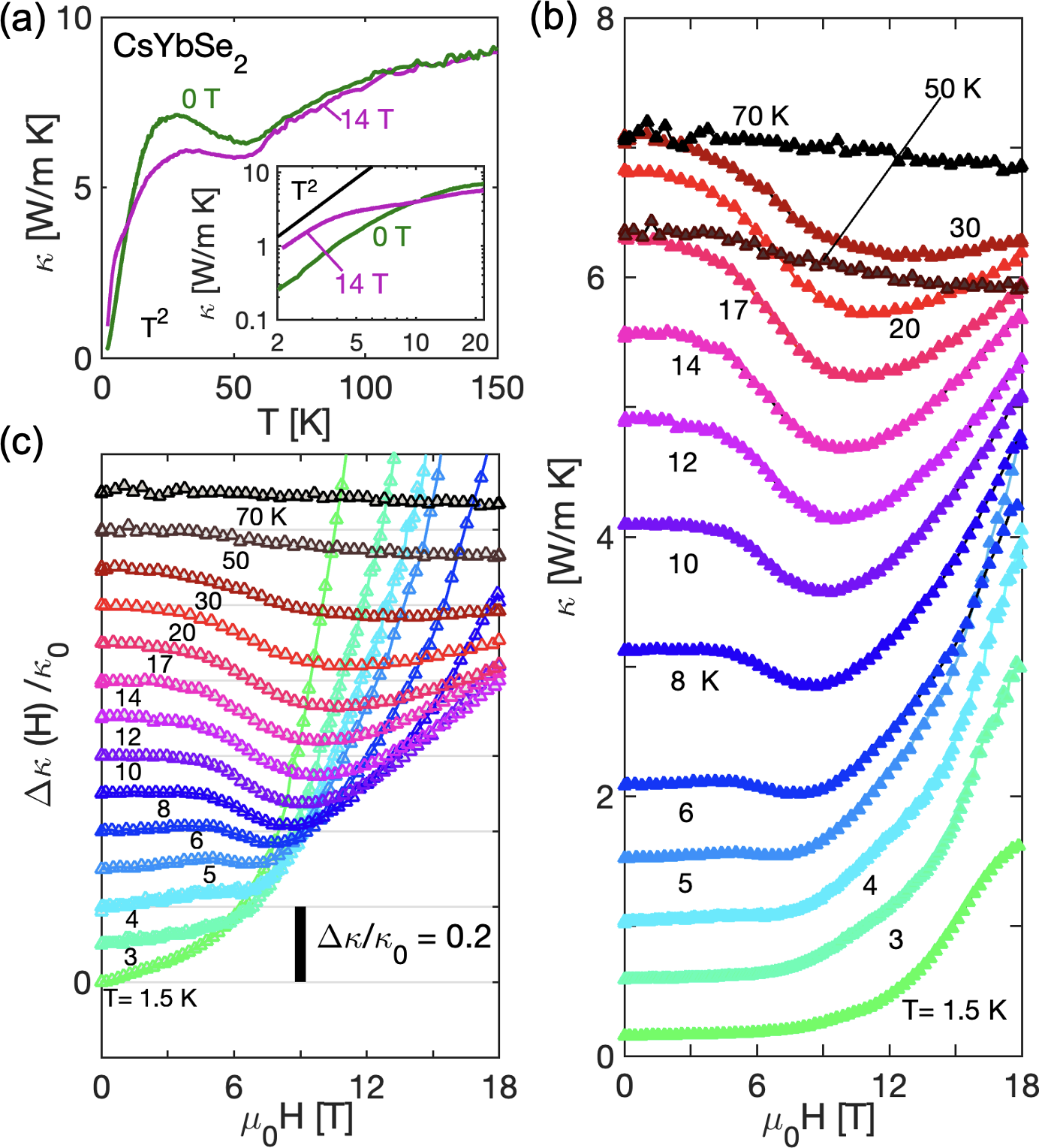}
\caption{(a) Temperature dependence  of thermal conductivity $\kappa$ at zero field  and at 14 T.
Inset shows the magnified view for $T<20$  K, where  $T^2$-like behavior can be seen at lower temperatures.
(b) Magnetic field dependence of $\kappa$ is shown at various temperatures. Above $T = 5$ K, a shallow minimum of $\kappa(H)$ appears at $H = \hmin$, where $\hmin$ increases with temperature, while for $T < 5$ K, $\kappa(H)$ is characterized by a rapid increase for $\muh > 6 $ T. 
(c)  Fractional thermal conductivity $\Delta \kappa(H)/\kappa_0$ is plotted as a function of $H$ at various temperatures with offset for clarity.  }
\label{kap_data}
\end{center}
\end{figure}
%%%%%%%%%%%%%%%%%%%%%%%%%%%%
%%%%%%%%%%%%%%%%%%%%%%%%%%%%

SFEs are excitations across the Zeeman gap, where a single pseudospin is flipped from its $S^x = -1/2$ ground state to the $S^x = +1/2$  excited state.  
%\red
SFEs are a kind of magnon excitation; we use the term SFE because it is more descriptive and specific for the spin excitations of our model.
\black
Without hybridization of SFEs and phonons, there is a flat band of non-propagating SFEs [Fig.~\ref{schem}(e)].  Hybridization modifies the dispersion of SFEs and phonons as shown schematically in Fig.~\ref{schem}(f), leading to two branches of mixed excitations with non-zero group velocities [Fig.~\ref{schem}(g)].   

The  \cybs~system  is particularly well-suited to explore the hybridization of acoustic phonons and SFEs:  (1) The CEF gap between the ground and first excited doublets
%\blue
$\Delta_{10} \approx 13$ meV  provides a large $T$ range
\black
where the effective spin-1/2 approximation is valid. (2) The small exchange energy ($J_{{\rm ex}} \approx$ 0.4 meV) combined with geometric frustration prevents long-range magnetic order 
%\blue
down to 0.3 K at zero field, with signs of field-induced local correlations below 1 K \cite{Xing2019, Pocs2021}, 
\black
leading to a wide paramagnetic regime where spin-spin exchange interactions may be neglected to a first approximation.

Fig.~\ref{kap_data} displays the thermal conductivity of \cybs~ as a function of temperature $1.5 \, \mathrm{K} <T< 150 \, \mathrm{K}$  and  field up to $\muh= 18$ T, within the paramagnetic state, where the temperature gradient and  field are parallel and within the $ab$ plane.
%\red
The monotonically increasing, field-independent behavior of $\kappa$ for $T \gtrsim 60$ K is not typical of crystalline materials, but similar behavior is commonly observed in non-magnetic amorphous solids \cite{Bermanbook, Bullen2000}.  We note that \cybs~and other delafossites form crystals in very thin layers that easily become separated along the $c$-axis, and are prone to stacking faults. We speculate this is responsible for the increasing thermal conductivity at higher temperatures.
\black
%remains $ \nabla T || H$ for the entire measurement.
For $T < 50$ K, the $T$-dependence of $\kappa$ is affected by field, 
where  $\mu_0 H = 14$ T enhances (suppresses) $\kappa$  at low (high) temperatures  as shown in Fig.~\ref{kap_data}(a).  The non-monotonic field dependence is clarified by plotting $\kappa$ versus $H$ at several values of $T$ in Fig.~\ref{kap_data}(b).  
When $T< 5$ K,  $\kappa(H)$ exhibits weak field-dependence as $H$ is first increased from zero, then rapidly increases upon further increasing $H$.  
As $T$ increases above  5 K,  $\kappa(H)$  shows a decrease with increasing $H$  until reaching a pronounced minimum at $H = \hmin$. The field $\hmin$ moves to large values as $T$ increases, eventually becoming hard to locate as the field-dependence is diminished.   
%The full $(T,H)$ region explored here fall within the paramagnetic state of \cybs. 
Fig.~\ref{kap_data}(c) show the fractional thermal conductivity, defined as  $\Delta \kappa(H)/\kappa_0 \equiv (\kappa(H) - \kappa_0)/\kappa_0$, where $\kappa_0$ is the value at zero field at a given $T$.  This quantity is plotted with a constant offset along the $y$-axis to highlight the monotonic increase of $\hmin$ with $T$.

To explain the  $\kap(H)$ data of  \cybs, we need to consider a coupling between phonons and magnetic excitations.  Recalling that the CEF effect arises from electrostatic interactions between a single magnetic ion and its surrounding ligands, small lattice distortions can create a modulation of the characteristic magnetic energy scales, leading to ME coupling.  
For example, varying the distortion of YbSe$_6$ octahedra will change the anisotropy of the $g$-tensor.
At sufficiently low $T$ when only the ground doublet is occupied, the most general linear ME coupling arising from local lattice distortions is $\mathcal{H}_{ME} = \mu_0 \mu_B \sum_{i, j} H_i \delta g_{i j}  \hat{S}^j$.  That is, ME coupling enters via a modulation of the $g$-tensor $\delta g_{i j}$. For small lattice distortions each component of $\delta g$ is a linear combination of components of the symmetric strain tensor $\epsilon_{i j} = \partial_i u_j + \partial_j u_i$, where $u_i$ is the displacement field.   We note that lattice distortions do not couple to the pseudospin in the limit of zero field, where time reversal symmetry holds and Kramers theorem prevents any splitting of the ground doublet.  
This contrasts with non-Kramers doublet systems such as TmVO$_4$, where the ME coupling remains non-zero in vanishing applied field \cite{Kjems1975}.

The most dramatic consequence of this ME coupling turns out to be the hybridization of SFEs and phonons, arising from terms corresponding to the emission/absorption processes illustrated in Fig.~\ref{schem}(c).
Here we introduce a highly simplified effective model designed to capture the essential qualitative physics of hybridization as manifested in thermal conductivity.  Our model can be motivated by a more microscopic treatment sketched the Appendix.  The SFEs are treated as bosons within a standard spin wave approximation, and we focus on only a single polarization of acoustic phonon.  The energies $E_{\pm}(k)$ of the two  branches of hybridized excitations are the eigenvalues of the matrix
%We devise a minimal models  with the ME coupling via the $g$-factor modulation with local displacement.  Using  the single-mode approximation of phonon for a simplicity,  the ME coupling Hamiltonian $\mathcal H_{\rm ME}$ is written  in  off-diagonal terms. 
\beq 
{\mathcal H}(k) = 
\Big (
\begin{array}{cc}
\hbar v_s k & \sqrt{\eta \hbar v_s k}  \gamma H \\
 \sqrt{\eta \hbar v_s k } \gamma H &\gamma H
\end{array}
\Big),
\label{MEmtx}
\eeq 
where $v_s$ is the sound velocity.
From specific heat in zero field (data not shown), we obtained  
$\hbar v_s = 12.97\, \mathrm{meV}$\AA ~equivalent to  $1.97\times10^3$ m/sec and also the Debye energy $\hbar \omega_D = 7.1 \, \mathrm{meV}$.  
 The off-diagonal matrix elements arise from ${\mathcal H}_{ME}$ with ME coupling strength parametrized by $\eta$.  
The $\sqrt{k}$-dependence matches the scaling of off-diagonal matrix elements within the treatment of the Appendix. Here, we neglect any angular dependence and assume the hybridized excitations have a spherically symmetric dispersion.  The off-diagonal terms in  Eq.~\ref{MEmtx} make it transparent that applied field simultaneously tunes the Zeeman gap $\gamma H$ and the strength of ME coupling.

%%%%%%%%%%%%%%%%%%%%%%%%%%%%
%%Figure  3%%%%%%%Figure 3 %%%%Figure 3%%
%%%%%%%%%%%%%%%%%%%%%%%%%%%%

 \begin{figure}[ht]
\begin{center}
\includegraphics[width= 1.0\linewidth]{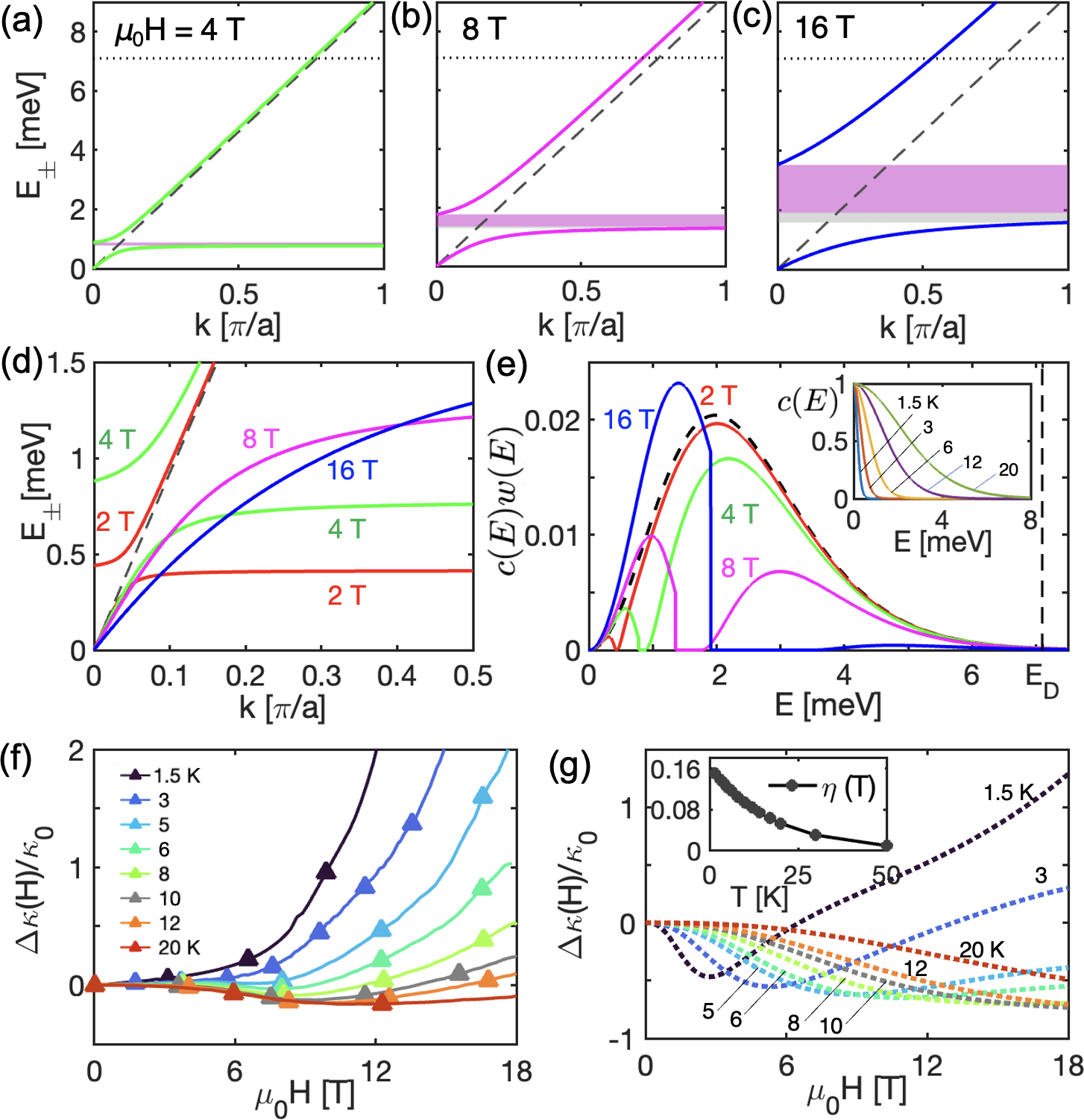}
\caption{ (a-c) $E_\pm (k) $ of Eq.~(\ref{eq:disp})  are  plotted  for  $\muh = 4,8 $ and 16 T, with $\eta = 0.13$ meV$^{-1}$.
Dashed lines show the phonon dispersion without ME coupling, while dotted horizontal lines indicate the Debye energy $E_D = 7.1$ meV. 
\black
 The energy gap $\Delta \propto H^2$ between upper and lower branches is shown with pink (integrating over all $k$)  and gray (cutoff at $k< \pi/a$) shading.     
(d) $E_{\pm}(k)$ for several values of applied field at smaller values of $k$.  
(e)  The integrand $c(E) w(E)$ of Eq. (\ref{kap_eq}) is shown for several values of applied field at $T = 6$ K. 
The upper and lower branches are clearly visible as two separate peaks separated by the gap, where $g(E) = 0$. 
The inset plots the specific heat $c(E)$ at selected temperatures.
%\red
Comparison of (f) measured and (g) calculated $\Delta\kappa/\kappa_0$ as a function of applied field at the temperatures as shown. Inset displays the empirical  $\eta(T)$. 
%with $\eta(T=1.5 K) = 0.15$ meV$^{-1}$. 
Despite the  highly simplified model, the calculations capture the essence of the data: non-monotonic field dependence with the minimum in $\kappa(H)$ at $\hmin$ moving to higher values as $T$ increases. \black
}
\label{fig3}
\end{center}
\end{figure}

The energies  of hybridized quasiparticles take the form
\begin{equation}
E_{\pm}(k) = \Big( \frac{ E_0 +  \gamma H}{2} \Big)
\pm \sqrt{ \Big( \frac{ E_0  -  \gamma H}{2} \Big)^2 +  \eta E_0 \gamma^2 H^2 } 
\label{eq:disp} \text{,}
\end{equation}
where $E_0 = \hbar v_s k$.  The dispersion of $E_{\pm}$ as a function of wavevector $k$  is plotted in Fig.~\ref{fig3}{\bf a}-{\bf c} for applied fields $\muh = 4,8,$ and 16 T, respectively.
\black
The lower branch $E_-(k)$ increases monotonically from zero to $\gamma H - \eta (\gamma H)^2$ as $k \to \infty$, while the upper branch $E_+(k)$ starts at  $E_+(k=0)= \gamma H$ and monotonically increases. 
The size of the gap $\Delta$ between the two branches thus takes the simple form $\Delta = \eta (\gamma H)^2$ 
when including states with arbitrarily large values of $k$.  %\blue   
The change in gap size  upon imposing the momentum cutoff $k<\pi/a$  is negligible, \black 
as illustrated by the gray shading in Fig.~\ref{fig3} (a-c); here, $a = 4.42$\AA~ is the $ab$-plane lattice constant.
%; this is shown in the gray shade in each panel.  
%The pink shading indicates the modestly increased gap size upon imposing the cutoff $k<\pi/a$. 
Note that the lower branch dispersion becomes flat with $E_-(k) = 0$ when $\eta = (\gamma H)^{-1}$, signaling an instability reached either at large ME coupling or strong applied field, beyond which the model will not be valid.  We thus always take $\eta < (\gamma H)^{-1}$, where $E_-(k)$ is non-zero and remains real.
Fig.~\ref{fig3}(d)  displays a magnified view of the low-energy region: 
 the slope (\emph{i.e.} group velocity) at $k=0$ decreases monotonically with increasing field. 
This is important for determining the high field behavior, as discussed below.

%\red 
SFEs are subject to a hard-core repulsive interaction that forbids two or more excitations from occupying the same lattice site.  This is neglected in a na\"{\i}ve linear spin-wave treatment, where SEFs are represented by bosonic particles for which arbitrarily high occupation number is allowed.  We account for the hard-core interaction at a mean-field level via temperature-dependent ME coupling $\eta = \eta(T)$, as described in the Appendix.  The effect of hard-core repulsion should become more pronounced with increasing temperature due to thermally excited SFEs, and as a result $\eta(T)$ is expected to decrease with increasing $T$.  In our calculations we employ an empirical monotonically decreasing form for $\eta(T)$, as shown in the inset of Fig.~\ref{fig3}(g).

Thermal conductivity is computed within a Debye-Callaway model \cite{Bermanbook} using the above hybridized quasiparticle spectra:
\begin{equation}
\kappa(T, H) = \sum_{\sigma = \pm} \int \frac{d \bk^3}{3(2\pi)^3} c(E_{\sigma}(\eta, k))v_{\sigma}^2(\eta, k) \tau(k) 
\label{kap_eq}. 
\end{equation}
Here, $v_{\pm} = (1/\hbar) d E_{\pm} / dk$ and $c(E) = E d n_B / dT$ is the specific heat of a single bosonic mode, with $n_B(T)$ the standard Bose occupation function, and we impose the momentum cutoff $k < \pi / a$. 
The scattering time $\tau(k)$  can be expressed $\tau^{-1} (k) = \sum \tau_i^{-1}$, where relaxation rates add for different scattering processes (\emph{e.g.} boundary, normal and Umklapp scattering), and each $\tau_i$ is a function of $k$ and/or $T$ \cite{Bermanbook}. 
%\red
Unlike pure phonons, a detailed theory for scattering of the hybrid phonon-SFE excitations is not available. To highlight the transport implications of hybridization alone, we start by considering the simplest case of $\tau = \tau_0$ independent of $k$, and return to the issue of potential $k$-dependence below.

With the $k$-independent relaxation time, we rewrite $\kappa$ as the energy integral 
\begin{equation}
\kappa(T, H) =  \frac{\tau_0}{3} \int_0^{\infty} dE \, c(E) w(E) ,
\label{kap_dos_eq}
\end{equation}
using the density of states $g(E)$ of the hybridized excitations, where we defined the spectral weight $w(E) = v^2(E) g(E)$.  
\black
In Fig.~\ref{fig3}(e), the integrand of Eq.~(\ref{kap_dos_eq}) is plotted  for a few values of $H$ at fixed $T = 6$ K. The upper and lower branches are clearly visible as two peaks separated by the gap.
As the field is increased from zero, the contribution of the upper branch decreases as spectral weight moves to higher energy, while that of the lower branch increases with field.
%\red
The evolution of $w(E)$ with applied field is shown in Fig. \ref{figA1} of the Appendix.
\black

The measured  $\fkap$ of \cybs~ and the calculated result from Eq.~(\ref{kap_dos_eq}) are shown, respectively, as a function of applied field in Fig.~\ref{fig3}(f) and~(g).
%\red [NEED to be replaced with $\eta (T)$ discussion]  In the calculation, we  use  $\eta = 0.13$ meV$^{-1}$,  only parameter here and take constant $\tau$ constant  as discucssed above and   independent of magnetic field, which is as appropriate for non-magnetic scattering.   \black
The temperature dependence of $\tau$ plays no role, as it cancels out in the fractional thermal conductivity; 
%\red 
again, we discuss $k$-dependence of $\tau$ below. \black
Our model qualitatively captures the essential characteristics of the data: non-monotonic field-dependence of $\kappa$ observed for $T > 5$ K, as well as the movement of $H_{{\rm min}}$ to larger values with increasing $T$.  
%\red
As should be expected for a highly simplified model, there are discrepancies between the data and the calculations, to which we return below.
\black

We now discuss how to understand the non-monotonic field dependence of $\kappa$ in terms of the hybridized phonon-SFE excitations based on simple  arguments, independent of the details of our model.  
We assume relatively weak ME coupling  such that  $\eta\gamma H < c$, where $c < 1$ is an arbitrary dimensionless constant chosen not too close to $1$;  a simple perturbation theory argument shows  that this condition prevents the hybridization from modifying the dispersion too strongly at large wave vector.  

First, we consider the low-$H$ regime where $\gamma H \ll k_B T$.  In general, excitations contribute more strongly to $\kappa$ when their energy density changes rapidly with temperature.  
This is dictated by the heat capacity $c(E)$ of bosonic excitations [inset of Fig.~\ref{fig3}(e)], 
%\red
which appears in the integrand
\black
of Eq.~\ref{kap_dos_eq}. $c(E)$ is only weakly $T$-dependent for $E \lesssim k_B T$ and decreases exponentially for $E \gtrsim k_B T$.  As in Fig. \ref{fig3}(a),  upon increasing $H$,  opening the gap between upper and low branches pushes spectral weight in the upper (lower) branch to higher (lower) energies.  
The effect of this on $\kappa$ is dominated by the upper branch, where $c(E$) falls off rapidly with energy, and pushing spectral weight to higher energies leads to a decrease in $\kappa$, as can be seen in
%\red 
Fig.~\ref{fig3}(e). \black
On the contrary,  for the energy scales in the lower branch, $c(E)$  only depends weakly on energy and hence the shift in spectral weight does not strongly affect $\kappa$.

Turning to the high-field $\gamma H \gg k_B T$ regime, only lower-branch states with $k \approx 0$ are appreciably thermally occupied, and the lower branch gives the dominant contribution to $\kappa$ and hence its field-dependence.
Moreover, we can approximate  the linear dispersion $E_-(k) \approx \hbar v_{{\rm eff}} k$ near $k = 0$ (Fig. ~\ref{fig3}(d)).  As $H$ increases, level repulsion bends the $k \approx 0$ lower-branch dispersion downward as illustrated in Fig.~\ref{fig3}(b), resulting in $v_{{\rm eff}}$ that decreases with increasing field.  Within our model, the lower-branch velocity at $k=0$ is indeed given by $v_- = v_s - v_s \eta \gamma H$.  
Examining Eq.~\ref{kap_dos_eq}, one might na\"{\i}vely conclude that $\kappa$ should decrease as $v_{{\rm eff}}$ decreases, given the factor of $v(E)^2 \approx v^2_{{\rm eff}}$ in the integrand. However, the density of states for a linearly dispersing mode of velocity $v_{{\rm eff}}$ is $g(E) = E^2 / 2 \pi^2 \hbar^3 v_{{\rm eff}}^3$, so in fact $\kappa \propto v^{-1}_{{\rm eff}}$ in the high-field regime, and $\kappa$ thus increases with increasing field, which is well-captured in Fig.~\ref{fig3}(g). 

%%%%%%%%%%%%%%%%%%%%%%%%%%%%
%%Figure  4  %%%%%%%Figure 4  %%%%Figure 4%%
%%%%%%%%%%%%%%%%%%%%%%%%%%%%

\begin{figure}[t]
\begin{center}
\includegraphics[width= 1.0\linewidth]{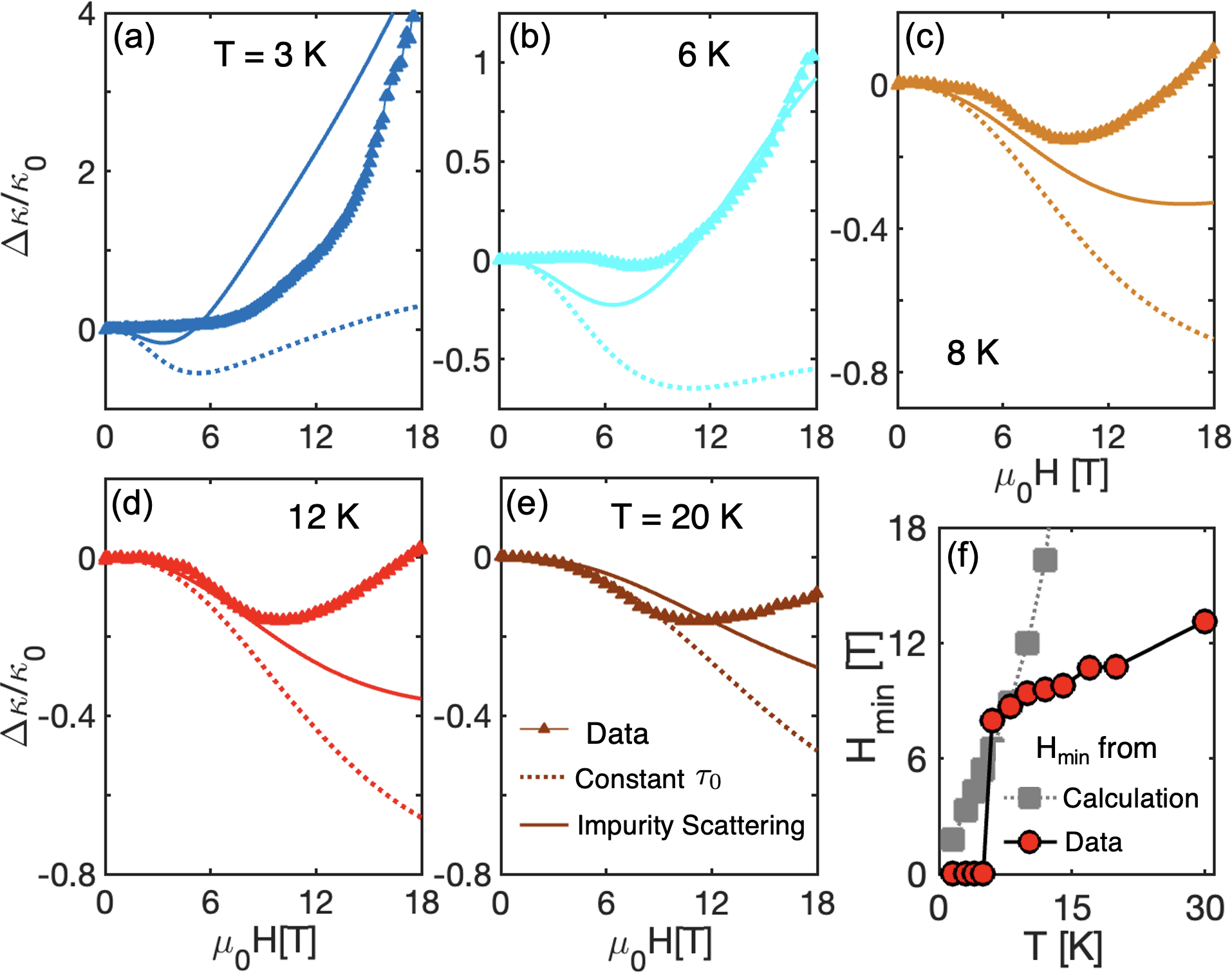}   
\caption{(a-e) ~Comparison  between measured (symbols) and calculated $\fkapH$  at different $T$ as shown.  
The dotted lines are from  Eq.~(\ref{kap_dos_eq}), the same results shown in  Fig.~\ref{fig3}(g). 
Solid lines are from Eq.~(\ref{eqn:kappa-imp}) with $v_b/v_s = 0.1$. 
(f) $\hmin$ obtained from the impurity scattering calculation (gray squares) and data (red circles) are plotted as a function of $T$. Lines are connecting the symbols.
%{\bf f} Calculated $\fkapH$ including $T$-dependent ME coupling strength $\eta = \eta(T)$, which incorporates the hard-core nature of SFEs at a mean-field level (see text and Methods).  $\eta(T)$ is chosen to optimize qualitative agreement with the data and is shown in the inset.  
%{\bf d} Calculated $\fkapH$, including scattering of phonons by paramagnetic fluctuations estimated using the deviation from saturation of the magnetization (see text), together with $\eta(T)$ in panel {\bf c}. 
%The relative strength of magnetic to non-magnetic scattering is quantified by $\alpha$ as described in the text. The temperature-dependence of $\alpha$ (inset) is chosen to optimize qualitative agreement with the data, where the monotonically decreasing $\alpha(T)$  is attributed to non-magnetic phonon-phonon scattering that increases with $T$.
% (a) $\fkap(H)$ data (solid line) and calculation using XXX  (broken line) are compared at selected $T$'s. (b) $\eta(T)$ correction was considered ($\alpha* n_{\rm mag}$ may com here
 %(a) First correction $\fkap(H)$ with the field dependent $\tau^{-1}(H) = \tau_0^{-1} (1+ \frac{M(H)-M_0}{M_0})$. (b) The  $\eta(T)$ incorporated in calculating the $\fkap(H)$ to account for  hard boson approximiation as shown in the inset.  Dashed line in (c-d) are  from the same data shown in (b) only for a fewer $T$'s
}
\label{kappa}
\end{center}
\end{figure}

%\red
Finally, we discuss the discrepancies between the $\fkap$ data and the calculation apparent in Fig.~\ref{fig3}(f) and (g). As compared to the calculation, the data shows a more pronounced increase of $\fkap$ at large applied field, and lacks the minimum in $\fkap(H)$ for  $T < 5 $ K.  
These differences prompt us to consider a simple phenomenological $k$-dependent scattering rate $\tau^{-1} =  \tau^{-1}_{0} + \tau^{-1}_{{\rm imp}}$, where $\tau_0$ is $k$-independent, and $\tau_{{\rm imp}} = \ell_{{\rm imp}} / v(k)$, as appropriate for scattering of hybrid phonon-SFE excitations off static impurities with a mean-free path $\ell_{{\rm imp}}$.  Writing $\tau^{-1}  = [ v(k) + v_b ] / \ell_{{\rm imp}}$, where we defined the velocity $v_b = \ell_{{\rm imp}} / \tau_0$, we have
 \begin{equation}
\kappa(T, H) = \frac{\ell_{{\rm imp}}}{3} \sum_{\sigma = \pm} \int \frac{d^3 \bk}{(2\pi)^3}  c(E_{\sigma}(k)) 
\frac{v_{\sigma}^2(k) }{ v_{\sigma}(k) + v_b }  \label{eqn:kappa-imp} \text{.}
% \\
%&=& 
%\frac{\ell_{{\rm imp}}}{3} \sum_{\sigma = \pm} \int \frac{d^3 \bk}{(2\pi)^3}  c(E_{\sigma}(k))  \frac{v_{\sigma}(k)}{1+v_b/v_{\sigma}(k)}  
\end{equation}
In the limit $v_b \gg v_{\sigma}(k)$, we recover our original model.  In the opposite limit $v_b \ll v_{\sigma}(k)$, the integrand is proportional to $c(E_{\sigma}(k)) v_{\sigma}(k)$ instead of $c(E_{\sigma}(k)) v_{\sigma}^2(k)$.  

Fig.~\ref{kappa}(a-e) displays $\fkap$  including impurity scattering according to Eq.~(\ref{eqn:kappa-imp}) together with the data and calculated $\fkap$ with $k$-independent $\tau_0$ [Eq. (\ref{kap_dos_eq})], at a few representative temperatures.  At low $T$, the qualitative agreement with the data is dramatically improved, with a shallower minimum, lower $\hmin$, and a steeper rise at high field.  The agreement at higher temperatures also improves, though discrepancies in detailed features still remain, as should be expected for this simplified and phenomenological model. Fig.~\ref{kappa}(f) shows the $T$-dependence of $\hmin$ obtained from the data and from $\fkap$ calculated using Eq.~(\ref{eqn:kappa-imp}). 
\black
 
%{\red  NEED to be removed  Second, we expect phonons to scatter off of fluctuating paramagnetic moments, with a scattering rate proportional to  the  effective density of the magnetic scattering centers $\nmag$, which is estimated as $\nmag (T, H) = \frac{\Delta M(T,H)}{M_S} = \frac{M(T,H)-M_S}{M_S}$, where $M_S$ is the saturated magnetization as treated in \cite{Pocs2020}.   Such scattering is thus suppressed with increasing applied field, as the magnetization approaches saturation.   This can be incorporated in a simple manner by replacing the scattering rate  with  $\tau^{-1}(H) =  \tau_0^{-1} \big(1+\alpha \nmag(H)\big)$ at a given $T$, where $\alpha$ depends on $T$ and measures the relative strength of magnetic to non-magnetic scattering.  Using the field dependence of magnetization calculated in the Weiss mean-field approximation \cite{Pocs2021} shown in Fig.~\ref{schem}{\bf b}, we replace the $\tau$ in Eq. (3) with $\tau(H)$  above to obtain $\fkap$ as a function of $H$ using the same  $\eta(T)$  in Fig.~\ref{kappa}{\bf c}  and $\alpha (T)$ displayed  in Fig.~\ref{kappa}{\bf d}. We find this correction has a much larger effect:  a strong rise in $\kappa$ at large field and significantly reduced size of the minimum $\fkap$ at $\hmin$, which further increases the qualitative agreement of the model with the data.  However, these corrections do not remove completely the model's pronounced minimum in $\kappa$ at low $T$ and small applied fields.  }

Needless to say, our minimal model does not account for material-specific details such as spin-spin exchange interactions, proximity to magnetic order,  multiple phonon polarizations, and anisotropy in the unperturbed phonon dispersion and ME coupling,  all of which may play a role.  We also expect that microscopic treatments of scattering processes of the hybridized quasiparticles will bring the model closer to the observed $\fkap$. 
%Moreover, a more detailed treatment both of the hybridized excitations and SEF-phonon scattering may be needed to capture the physics within this regime.

In summary,  our study demonstrated that the field dependence of thermal conductivity in \cybs \, can be 
attributed to heat transport by hybridized quasiparticles formed from acoustic phonons and SFEs. 
Our highly simplified model qualitatively captures (1) the initial decrease of $\kappa$ under applied magnetic field to a minimum at $H = \hmin$, followed by an increase at higher fields and (2) the monotonic increase of $\hmin$ with $T$.  
%\red
Including impurity scattering via a $k$-dependent relaxation time improves the qualitative agreement between the model and the data.
\black
The key ingredients of our model are Zeeman splitting, acoustic phonons, and weak ME coupling via modulation of the magnetic $g$-tensor by local strain, all of which are found in many systems. 
We thus expect that phonon-SFE hybridization will  be essential as a  starting point to understand the field dependence of  thermal transport in a wide range of magnetic insulators.

\begin{acknowledgments}
We thank Y. Matsuda and  Y. Kasahara for helpful discussions. 
Work at University of Colorado Boulder was supported by the U.S. Department of Energy, Office of Science, Basic Energy Sciences (BES) under Award No. DE-SC0021377 (experimental work by C.A.P. I.A.L. and M.L.) and Award No. DE-SC0014415 (theoretical work by M.H.).
Work at Oak Ridge National 
Laboratory (ORNL) was supported by the U.S. Department of Energy, Office of Science, Basic Energy Sciences, Materials Sciences and Engineering Division.
A portion of this work was performed at the National High Magnetic Field 
Laboratory, which is supported by the National Science Foundation Cooperative 
Agreement No. DMR-1644779, the State of Florida, and the U.S. Department of Energy. 
\end{acknowledgments}

\appendix*

\renewcommand{\thefigure}{A\arabic{figure}}

\setcounter{figure}{0}
\section{} 
\subsection {Microscopic Theoretical Treatment of Magnetoelastic Coupling} 
Here we sketch a more microscopic theoretical treatment of the problem of coupled acoustic phonons and SFEs, which motivates the simplified effective model described in the main text.  We treat the acoustic phonons as excitations of a continuous elastic medium with displacement field $u_i(\br)$ and symmetric stress tensor $\epsilon_{i j} = \partial_i u_j + \partial_j u_i$.  Spin-1/2 spins $\hat{S}^i_{\br}$ lie on the sites of a Bravais lattice, and the ME coupling takes the form ${\cal H}_{ME} = \sum_{\br} {\cal H}_{ME}(\br)$, where the sum is over lattice sites and
${\cal H}_{ME}(\br) = \mu_0 \mu_B H_x \delta g_{x i}(\br) \hat{S}^i_{\br}$
as described in the main text.  (In this Appendix, sums over repeated indices are implied.)  We drop the $i=x$ term as it does not lead to hybridization of phonons and SFEs, and the remaining terms can be written ${\cal H}_{ME}(\br) = \mu_0 \mu_B H_x [ \Lambda_{i j} \epsilon_{i j} \hat{S}^+_{\br} + \text{H.c.}]$, where $\Lambda_{i j}$ is a complex matrix of coupling constants parametrizing the ME coupling and the spin raising operators are defined by $\hat{S}^+_{\br} \equiv \hat{S}^y_{\br} + i \hat{S}^z_{\br}$ (also $S^-_{\br} = (S^+_{\br})^\dagger$ for spin lowering operators).  These spin raising/lowering operators are defined to raise/lower $\hat{S}^x_{\br}$, corresponding to the direction of applied field.

To study the effect of ${\cal H}_{ME}$, we go to momentum space.  The Fourier transform of the diplacement field is
\begin{equation}
u_i(\br) = \sum_{\bk, \lambda} \sqrt{ \frac{\hbar}{2 V \rho \omega_\lambda(\bk) } } e^{i \bk \cdot \br} \hat{e}_{\lambda i}(\bk) [a^{\vphantom\dagger}_{\bk \lambda} + a^\dagger_{\bk \lambda} ] \text{,}
\end{equation}
where $V$ is the volume, $\rho$ the mass density of the crystal, and $\lambda$ labels the three phonon polarizations with frequencies $\omega_\lambda(\bk) = v_\lambda k$, polarization vectors $\hat{e}_\lambda(\bk)$, and creation operators $a^\dagger_{\bk \lambda}$.  Because we treat the lattice as a continuous medium, the magnitude of the wave vector $\bk$ is unrestricted in the sum.  For the spin operators, we define the Fourier transform by
\begin{equation}
S^+_{\br} = \frac{1}{\sqrt{N}} \sum_{\bk \in \mathrm{BZ}} e^{-i \bk \cdot \br} S^+_{\bk} \text{,}  \label{eqn:sft}
\end{equation}
where $N$ is the number of lattice sites and the wave vector sum is restricted to the first Brillouin zone.  

To make a linear spin-wave approximation, we introduce Holstein-Primakoff bosons with creation operators $b^\dagger_{\br}$ by writing $S^+_{\br} = b^\dagger_{\br} \sqrt{1 - b^\dagger_{\br} b^{\vphantom\dagger}_{\br} }$.  We note the form of Eq.~(\ref{eqn:sft}) is chosen so that if we ignore the square root and thus replace $S^+_{\br}$ by $b^\dagger_{\br}$ and $S^+_{\bk}$ by $b^\dagger_{\bk}$, the momentum space creation/annihilation operators satisfy canonical commutation relations.  Simply dropping the square root neglects the hard-core nature of the Holstein-Primakoff bosons, but we can restore this effect at a mean-field level by replacing $S^+_{\br} \to \sqrt{1 - \bar{n} } b^\dagger_{\br}$, where $\bar{n} = \langle b^\dagger_{\br} b^{\vphantom\dagger}_{\br} \rangle$.  

Plugging the Fourier transforms into ${\cal H}_{ME}$, keeping only hybridization terms proportional to $a^\dagger_{\bk \lambda} b^{\vphantom\dagger}_{\bk}$ (or the Hermitian conjugate), and dropping contributions from higher energy phonons outside the first Brillouin zone, we have
\begin{widetext}
\begin{equation}
{\cal H}_{ME} = \sqrt{1 - \bar{n}} \mu_0 \mu_B H_x \sum_{\bk \in \mathrm{BZ}} \sum_{\lambda}  \Big\{ \sqrt{\frac{\hbar}{2 v_{uc} \rho \omega_\lambda(\bk) } } \Lambda_{i j} [i k_i \hat{e}_{\lambda j}(\bk) + i k_j \hat{e}_{\lambda i}(\bk) ] a^{\vphantom\dagger}_{\bk \lambda} b^\dagger_{\bk} + \text{H.c.} \Big\}  \text{.}  \label{eqn:hme-micro}
\end{equation}
\end{widetext}
Here $v_{uc}$ is the volume of a crystalline unit cell.  We note that the matrix elements of this Hamiltonian are proportional to $\sqrt{k}$, which comes from the factors of $k_i / \sqrt{\omega_\lambda(\bk)}$; this motivates the $\sqrt{k}$-dependence of the off-diagonal matrix elements in the effective model of the main text.  Moreover, the effect of the mean-field correction $\sqrt{1- \bar{n}}$ is to renormalize the overall strength of the ME coupling, giving a temperature-dependent coupling that goes down as thermal occupation of SFEs increases.

Finally, we remark that it would be possible to compute $\kappa$ along the same lines as in the main text using the ME coupling of Eq.~(\ref{eqn:hme-micro}) and including all three phonon polarizations.  While this may be valuable to explore in future work, it is important to emphasize that this introduces additional adjustable parameters associated with couplings to different phonon polarizations, significantly increasing the complexity of the model.

\subsection{Spectral weight function}
%%%%%%%%%%%%%%%%%%%%%%%%%%%
%Figure A1 %%%%%Figure A1%%%%Figure A1%%
%%%%%%%%%%%%%%%%%%%%%%%%%%%
\begin{figure}%[!b]%[ht]
\begin{center}
\includegraphics[width= 1.0\linewidth]{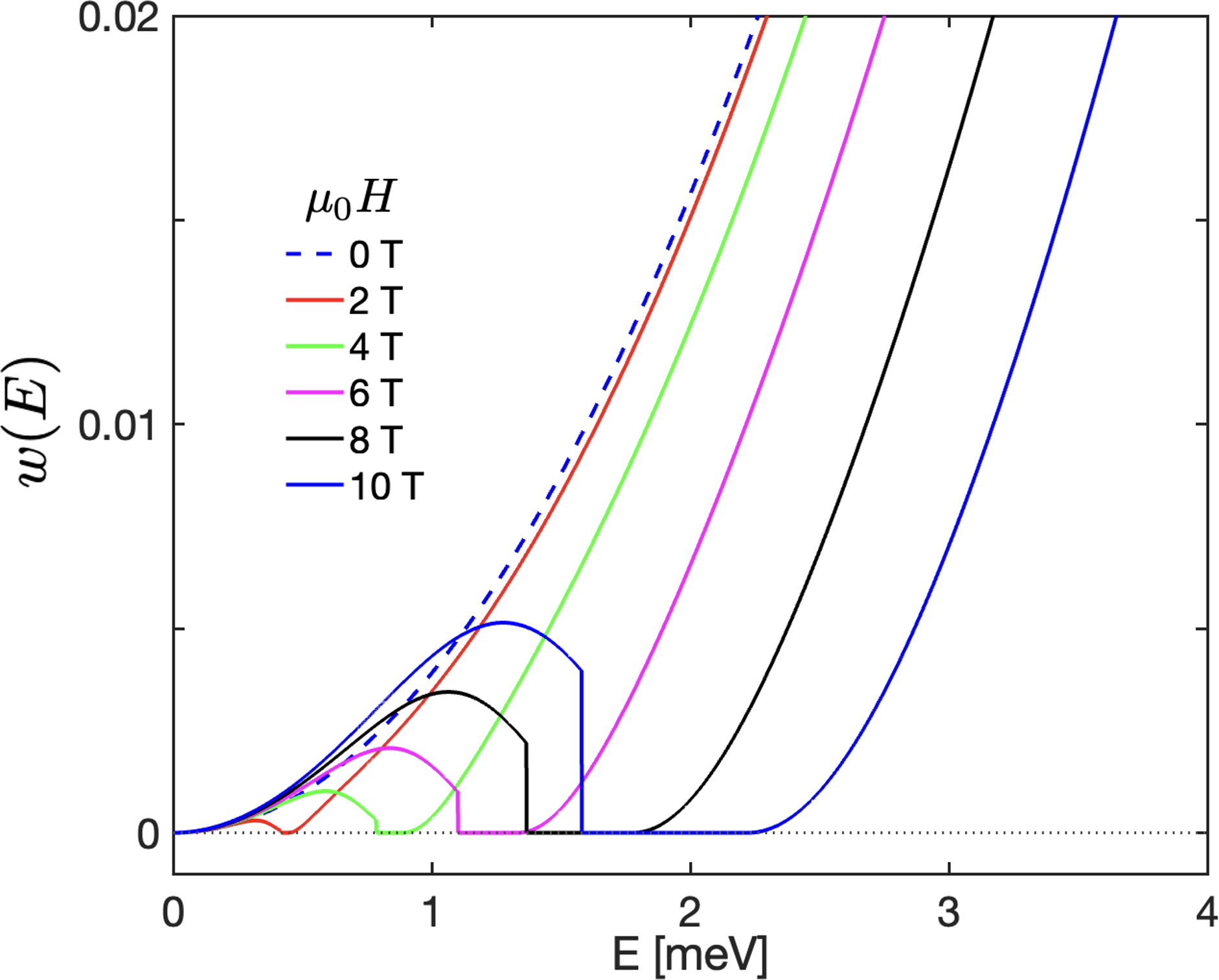}
\caption{The evolution of the spectral weight $w(E) = g(E) v^2(E)$ as a function of magnetic field.}
\label{figA1}
\end{center}
\end{figure}
Fig.~\ref{figA1} plots the spectral weight function  $w(E) = g(E)v^2(E)$  as a function of $E$  at several different magnetic field values as shown.  Within the gap between upper and lower branches, $w(E) = 0$.  With increasing magnetic field, the upper branch spectral weight shifts to higher energies, while the lower branch spectral weight has a peak that grows and moves to higher energies.

%\subsection{Samples and Experimental }
%Millimeter-sized hexagonal shape CsYbSe$_2$ single crystals were grown by salt flux method
%following the procedure described in Ref. \cite{Xing2020acs}.
%The in-plane longitudinal thermal conductivity was measured on the samples of typical dimensions $1.5 \times 3\times 0.2$ mm$^3$ using a single-heater, two-thermometer configuration in steady-state operation with the field applied in the $ab$ plane and in the direction of the thermal gradient. All thermometry was performed using Cernox resistors, which were precalibrated individually and \emph{in situ} under the maximum applied fields of each instrument.

%
\bibliographystyle{apsrev4-2}
%\bibliography{cyskappa_ref.bib}


\end{document}